\documentclass[12pt]{article}
\usepackage{alphabeta}
 \usepackage{amssymb}
 \usepackage{amsmath}
  \usepackage{graphicx}
 \usepackage[utf8]{inputenc}
 \usepackage{mathcomp}
 \usepackage{mathtools}
\usepackage{mdwlist}
\usepackage{bm}
\usepackage{enumerate}
\usepackage{times}
\usepackage[labelsep = space,font={small},labelfont={sf,bf},textfont=sf]{caption}
\usepackage[a4paper, total={7.086in, 9.1in}]{geometry}
\usepackage{setspace}

\makeatletter
\renewcommand{\maketitle}{\newpage\setlength{\parskip}{12pt}
    {\Large\bfseries\noindent\sloppy\textsf{\@title} \par}%
    {\noindent\sloppy \@author}%
}

\renewcommand{\section}{\@startsection {section}{1}{0pt}%
    {-6pt}{1pt}%
    {\bfseries}%
    }
\renewcommand{\subsection}{\@startsection {subsection}{2}{0pt}%
    {-0pt}{-0.5em}%
    {\bfseries}*%
    }

\renewenvironment{figure}{\let\caption\NAT@figcaption}{}
    
\newcommand{\NAT@figcaption}[2][]{{%
    \refstepcounter{figure}{
        \par
    }
    \sffamily\noindent\textbf{Figure \arabic{figure}}\hspace{0.5em}#2}
    }
    
\makeatother

\newenvironment{affiliations}{%
    \setcounter{enumi}{1}%
    \setlength{\parindent}{0in}%
    \slshape\sloppy%
    \begin{list}{\upshape$^{\arabic{enumi}}$}{%
        \usecounter{enumi}%
        \setlength{\leftmargin}{0in}%
        \setlength{\topsep}{0in}%
        \setlength{\labelsep}{0in}%
        \setlength{\labelwidth}{0in}%
        \setlength{\listparindent}{0in}%
        \setlength{\itemsep}{0ex}%
        \setlength{\parsep}{0in}%
        }
    }{\end{list}\par\vspace{-13pt}}

\renewenvironment{abstract}{%
    \setlength{\parindent}{0in}%
    \setlength{\parskip}{0in}%
    \bfseries%
    }{\par\vspace{-6pt}}
    
\title{{{Magnetic Excitations in Non-Collinear Antiferromagnetic Weyl Semimetal \boldmath\({\mathsf{Mn_{3}Sn}}\)}} }

\author{\parbox{\linewidth}{\setlength{\baselineskip}{7mm}Pyeongjae Park$^{1,2}$, Joosung Oh$^{1,2}$, Klára Uhlířová$^{3}$, Jerome Jackson$^{4}$, András Deák$^{5,6}$,  László Szunyogh$^{5,6}$, Ki Hoon Lee$^{1,2}$, Hwanbeom Cho$^{1,2}$, Ha-Leem Kim$^{2}$, Helen C. Walker$^{7}$, Devashibhai Adroja$^{7,8}$, Vladimír Sechovský$^{3}$ \& Je-Geun Park$^{1,2*}$ }}
\date{}
\begin{document}

\setlength{\parskip}{0.5em}
\setlength{\parindent}{1em}
\maketitle
\setlength{\baselineskip}{7mm}

\begin{affiliations}
 \item Center for Correlated Electron Systems, Institute for Basic Science, Seoul 08826, Korea
 \item Department of Physics and Astronomy, Seoul National University, Kwanak-Gu, Seoul 08826, Korea
 \item Charles University in Prague, Faculty of Mathematics and Physics, Ke Karlovu 5, 121 16 Prague 2, Czech Republic
 \item Scientific Computing Department, STFC Daresbury Laboratory, Warrington, WA4 4AD, United Kingdom
 \item Department of Theoretical Physics, Budapest University of Technology and Economics, Budafokiút 8. H-1111 Budapest, Hungary
 \item 2MTA-BME Condensed Matter Research Group, Budafoki út 8, H-1111 Budapest, Hungary
 \item ISIS Facility, Rutherford Appleton Laboratory, Didcot OX11 0QX, UK
 ISIS Facility, Rutherford Appleton Laboratory, Chilton, Didcot, Oxon, OX11 0QX, United Kingdom
 \item Highly Correlated Matter Research Group, Physics Department, University of Johannesburg, P.O. Box 524, Auckland Park 2006, South Africa
\end{affiliations}

 \setlength{\parindent}{0em}
* : jgpark10@snu.ac.kr
\setlength{\parindent}{1em}
 \vskip 0.3in

\begin{abstract}
\textsf{\boldmath\(\mathsf{Mn_{3}Sn}\) has recently attracted considerable attention as a magnetic Weyl semimetal exhibiting concomitant transport anomalies at room temperature. The topology of the electronic bands, their relation to the magnetic ground state and their nonzero Berry curvature lie at the heart of the problem. The examination of the full magnetic Hamiltonian reveals otherwise hidden aspects of these unusual physical properties. Here, we report the full spin wave spectra of \(\mathsf{Mn_{3}Sn}\) measured over a wide momentum - energy range by the inelastic neutron scattering technique. Using a linear spin wave theory, we determine a suitable magnetic Hamiltonian which not only explains the experimental results but also stabilizes the low-temperature helical phase, consistent with our DFT calculations. The effect of this helical ordering on topological band structures is further examined using a tight-binding method, which confirms the elimination of Weyl points in the helical phase. Our work provides a rare example of the intimate coupling between the electronic and spin degrees of freedom for a magnetic Weyl semimetal system.}
\end{abstract}
\newpage

\doublespacing
\normalsize
\section*{Introduction}
It is an intriguing question how certain electronic band structures give rise to nontrivial topological properties. Initial work focused on how unique edge states form for particular symmetries, leading to the now broad and popular field of topological insulators\(\mathrm{^{1}}\). More recently, attention has concentrated on magnetic systems, where Dirac points at band crossings are broken into two sets of so-called Weyl points because of the broken time-reversal symmetry. Systems with Weyl points exhibit a nontrivial Berry’s phase, which can be characterized by measuring transport anomalies such as the anomalous Hall effect (AHE). \(\mathrm{Mn_{3}Sn}\), a non-collinear metallic antiferromagnet whose magnetic ions sit on a kagome lattice, has recently drawn growing interest for its rather remarkable transport anomalies. Density functional calculations examining the Weyl points\(\mathrm{^{2-3}}\), and transport measurements\(\mathrm{^{4-8}}\) of \(\mathrm{Mn_{3}Sn}\) show that this antiferromagnetic metal exhibits significant transport anomalies at room temperature. Subsequently, the presence of the Weyl points in the electronic band structure have been confirmed by angle-resolved photoemission spectroscopy (ARPES) and magnetoresistance measurements\(\mathrm{^{9}}\). More recently, it was suggested that it may also have spin polarized current, which would make it an interesting candidate for spintronics applications\(\mathrm{^{10}}\).

According to studies done in the 1980's and 1990's the magnetic structure of \(\mathrm{Mn_{3}Sn}\) has a \(120^\circ \) structure with a negative vector chirality at room temperature. This ground state indicates the presence of significant Dzyaloshinskii-Moriya (DM) interactions mediated by a spin-orbit coupling\(\mathrm{^{11-14}}\), which, together with an easy-axis anisotropy, stabilizes a weak ferromagnetic moment\(\mathrm{^{13,15}}\). On the other hand, the low temperature phase of \(\mathrm{Mn_{3}Sn}\) is known to depend on the precise ratio between Mn and Sn contents (ref. 16). Some earlier results show that \(\mathrm{Mn_{3}Sn}\) displays an additional incommensurate helical ordering along the c-axis \(\mathrm{(k_{z}\simeq 0.09)}\) below 220\(\mathrm{\sim}\)270 K\(\mathrm{^{17,18}}\) (we call it as the A-type \(\mathrm{Mn_{3}Sn}\)), while others found a glass phase below 50 K, instead of the helical ordering\(\mathrm{^{19,20}}\) (we call it as the B-type \(\mathrm{Mn_{3}Sn}\)). While both types show the same transport anomalies at room temperature\(\mathrm{^{4,7}}\), the A-type \(\mathrm{Mn_{3}Sn}\) does not show the transport anomaly at the low temperature phase with the helical ordering\(\mathrm{^{21}}\).

It is usually beneficial to carry out inelastic neutron scattering experiments at lower temperature because of thermal fluctuations. However, the complex glass phase of the B-type \(\mathrm{Mn_{3}Sn}\) makes it difficult to use the inelastic neutron scattering technique at low temperature. Meanwhile, we note that both types are identical in terms of the crystal and magnetic structures at room temperature\(\mathrm{^{9,18,20,21}}\), and it is also known that the phase transition at 220\(\mathrm{\sim}\)270 K in the A-type \(\mathrm{Mn_{3}Sn}\) is not accompanied by any structural transition\(\mathrm{^{22}}\). Thus, we can gain insight into the magnetic Hamiltonian of \(\mathrm{Mn_{3}Sn}\) at room temperature by measuring the A-type \(\mathrm{Mn_{3}Sn}\) at low temperature, since there should be, a priori, no significant difference in the intrinsic magnetic Hamiltonian for both types of \(\mathrm{Mn_{3}Sn}\) samples with the same crystal structure.

The magnetic excitation spectra of the A-type, helically ordered, \(\mathrm{Mn_{3}Sn}\) were previously studied using inelastic neutron scattering (INS) in the early 1990’s with limited success\(\mathrm{^{23-25}}\). Here, we report the full spin wave spectra of the A-type \(\mathrm{Mn_{3}Sn}\) over a wide momentum-energy range measured by inelastic neutron scattering. The full magnetic Hamiltonian based on a local moment model is proposed in order to explain the data completely using a linear spin wave theory. Supported by density functional theory (DFT) calculations, we also present a low temperature magnetic structure which involves a helical ordering with some important details that have so far been neglected in previous studies, including its effect on the Weyl points of \(\mathrm{Mn_{3}Sn}\).

\section*{Results}
\subsection{Magnetic Structure}
A \(120^\circ \) magnetic structure with negative vector chirality (\(\mathrm{\Gamma_{5}}\)) is used within each single a-b plane, which has been confirmed by both theoretical calculations and experiments\(\mathrm{^{11-15}}\) (Fig 1b). However, because our inelastic neutron scattering experiment was done at 5 K, extra effects from the low temperature phase must be considered rather than just using the room temperature phase. As we noted in the introduction, there are two types of \(\mathrm{Mn_{3}Sn}\)\(\colon\) A and B-type. By measuring the magnetization of our sample, we observed a clear phase transition near 260 K but failed to observe any increasing magnetization below 50 K (see Supplementary Fig. 1). This result confirms that our sample has the helical order of the A-type at low temperature as shown in Fig 1a.

There are a few other noteworthy features of this phase. First, the spin directions of two Mn atoms connected by inversion symmetry are no longer parallel in the helical phase. Instead, one of them is rotated by as much as half a turn (\(\simeq16.3^\circ\)) in the a-b plane by the helical ordering (Fig. 1a). This helical ordering has significant effects on Weyl fermions and the corresponding Hall conductivity of \(\mathrm{Mn_{3}Sn}\), which will be dealt with in the discussion section. The helical ordering of the A-type \(\mathrm{Mn_{3}Sn}\) is rather complicated\(\colon\) a previous neutron diffraction result shows that there are multiple satellite peaks at \((1,0,\tau_{1})\), \((1,0,\tau_{2})\), and \((1,0,3\tau_{2})\), with \(\tau_{1}\simeq0.07\) and \(\tau_{2}\simeq0.09\) (ref. 17). The first two peaks correspond to two propagating helices, and the third one is due to the anharmonicity of the second helix. We can safely ignore the contribution by \(\tau_{1}\) for our spin waves calculations, since the diffraction peak of \(\tau_{1}\) is much weaker than the other one\(\mathrm{^{26}}\). The existence of anharmonicity can also be well explained by our Hamiltonian, which will be shown later.

\subsection{INS data and magnetic Hamiltonian}
Our inelastic neutron scattering data show the spin wave spectra over the full Brillouin zone, which extends to almost 0.1 eV (Fig 2a). These coherent spin wave spectra prove to have a well-defined magnetic structure at 5 K. To explain the data, we use the following Hamiltonian\(\colon\)
\begin{equation}
\hat{H}= \sum_{i,k}^{} {(J_{k}\mathbf{S}_{i}\cdot \mathbf{S}_{i+\Delta k})} + \sum_{i,k}^{} {\mathbf{D}_{k}\cdot(\mathbf{S}_{i}\times \mathbf{S}_{i+\Delta k})} \\
+ \sum_{i}^{} {K(\hat{n_i}\cdot \mathbf{S}_{i})^2}+ B^{6}_{6}\sum_{i}^{} {{Q_{i}}^{6}_{6}}
\end{equation}
where \(\Delta k\) is the bond vector corresponding to the \(\mathrm{k^{th}}\) nearest neighbor. The first two terms denote the usual isotropic exchange interaction and the DM interaction between Mn atom and its \(\mathrm{k^{th}}\) nearest neighbor, while the last two terms denote a single ion anisotropy related to the local easy axes on the a-b plane and the \(\mathrm{6^{th}}\) order anisotropy term, respectively. The last term originates from crystal field effects, described in terms of Stevens operator equivalents\(\mathrm{^{27}}\). The necessity of this last term will be discussed in further detail later in this paper. With this Hamiltonian, we fitted the measured dispersion curves using a linear spin wave theory. Indicated by solid lines layered on Fig 2a, the fitted result from the Hamiltonian with coupling constants stated in Table 1 is consistent with the data. For our fitting, we used seven isotropic exchange parameters\(\colon\)up to the \(\mathrm{2^{nd}}\) in-plane nearest neighbor coupling and up to the \(\mathrm{3^{rd}}\) inter-plane nearest neighbor coupling.

We have also taken into account the effects of magnon damping and twins for our spin wave calculations. The calculated spin wave spectra from a linear spin theory yield very large spectral weight above 80 meV (See Supplementary Fig. 3), whereas our data show quite a uniform intensity throughout almost the whole energy range. To resolve this discrepancy, energy-dependent magnon damping is included to the calculation, which is often necessary to explain the spin waves for metallic magnets\(\mathrm{^{28,29}}\) (see Supplementary Note 3 for theoretical background). The damping effect we considered is similar to that of refs 28, 29. By analyzing the elastic peak positions of the INS data, we found two kinds of twins rotated about \(\mathrm{\pm 3.5^{\circ}} \) from the original lattice orientation. The calculated spin wave spectra with both twins and damping effect included are quite consistent with the data, supporting our Hamiltonian (Fig 2b \& Fig 3).

When fitting the magnetic Hamiltonian, we also performed Monte Carlo simulations before the spin wave calculations to check whether our Hamiltonian stabilizes the magnetic structure mentioned before. Indeed, our Hamiltonian successfully demonstrates a helical ordering as the ground state, with \(k_{z}\) of 0.0904. This stabilization is found in our model calculations to arise from exchange frustration, particularly due to the positive \(\mathrm{J_{6}}\) and the negative \(\mathrm{J_{4}}\) and \(\mathrm{J_{5}}\). Note that the optimized \(k_{z}\) value is found to be narrowly focused around 0.0904\(\pm 0.01\), since the resulting \(k_{z}\) value varies sharply even if the J values are adjusted only slightly.

\subsection{Spin-model based on DFT calculations}
In order to render further credence to the above study of the helical ordering in \(\mathrm{Mn_{3}Sn}\) in terms of equation (1), we performed ab-initio calculations of the exchange interactions. Indeed, the dominating interactions fall within a distance of 5-6 {\AA} (see Fig. 4a) as predicted by our model Hamiltonian (Table 1). Though the corresponding values of the isotropic interactions from the fitting to the magnon spectrum and from ab-initio calculations differs, there are crucial similarities between the two sets of parameters: (i) strong in-plane antiferromagnetic interactions that stabilize the triangular in-plane spin-structure and (ii) the above mentioned frustration of out-of-plane exchange interactions being the source of the helical modulation. This is evidenced in Fig 4b showing the dispersion \(\mathrm{E(k_{z})}\) of the helically modulated \(\mathrm{\Gamma_{5}}\) spin-state of \(\mathrm{Mn_{3}Sn}\) as calculated with the ab-initio exchange interactions: a minimum of \(\mathrm{E(k_{z})}\) emerges at \(\mathrm{k_{z}=0.086 (2\pi/c)}\) (blue curve in Fig 4b), in very good agreement with the desired value of 0.0904.

\subsection{DM interaction and single ion anisotropy}
As shown in Table 1, we applied the DM interaction on the nearest in-plane neighbor coupling (\(\mathrm{J_{2}}\) and \(\mathrm{J_{3}}\)) with its DM vector being parallel to the \( \hat{z} \) axis (with the direction from site i to site j defined counterclockwise in the triangle formed by \(\mathrm{J_{2}}\)). The most important role of this term is to stabilize the magnetic structure with negative vector chirality (\(\mathrm{\Gamma_{5}}\)) since this term lowers the energy of that structure\(\mathrm{^{12-14}}\). It also serves as an effective easy-plane anisotropy, stabilizing the structure in which the spins lie in the a-b plane. For these reasons, we assumed that the easy-plane anisotropy effect purely comes from our DM interaction term, rather than a single-ion anisotropy term with its easy plane perpendicular to the c-axis, similar to a recent theoretical study\(\mathrm{^{30}}\).

To explore the effect of DM interactions more quantitatively, we calculated the DM interactions in the system using relativistic DFT\(\mathrm{^{31}}\). The dominating z-components of the DM vectors are smaller at least by two orders of magnitude than the leading isotropic interactions (Fig 4a), consistent with our fitting results (see Table 1). When including the DM interactions to the calculation of \(\mathrm{E(k_{z})}\), we observe that the curve is shifted downwards (at \(\mathrm{k_{z}=0}\) by about 3 meV) stabilizing the \(\mathrm{\Gamma_{5}}\) spin-state against the \(\mathrm{\Gamma_{3}}\) state with different chirality (red line in Fig 4b). As the DM energy \(\mathrm{E_{DM}(k_{z})}\) monotonously increases with \(\mathrm{k_{z}}\) (inset of Fig 4b), the out-of-plane DM interactions somewhat destabilize the helical ordering but this effect is marginal. Surprisingly, \(\mathrm{E_{DM}(k_{z})}\) shows a quadratic dependence for small \(\mathrm{k_{z}}\), in contrast to the usual linear dependence of spin-spirals with a ferromagnetic order normal to the propagation. Analytic calculations with nearest neighbor out-of-plane DM vectors indeed show that \(\mathrm{E_{DM}(k_{z})\simeq cos(k_{z}c/2})\), which is fully confirmed numerically (inset of Fig 4b).

When it comes to an analysis of the DM interaction in the data, the low energy dynamics of the spin wave is important since the DM interaction is relatively small and has little influence on the spectra in Fig 2 (ref. 32). Figure 5 shows the low energy INS data obtained from the incident neutron energies relatively lower than in Fig 2, and the corresponding simulation results. The size of DM interaction is determined through implementing a gradually increasing mode at an energy value around 16 meV (see inset of Fig 5d, indicated as a red arrow near 16 meV) which is also seen in the INS data of the previous study (ref. 25). Meanwhile, the result from our relativistic DFT gives the magnitude of the in-plane nearest neighbor DM interaction as 0.17 meV, the same order of magnitude as the fitted values.

We also found single ion anisotropy to be as crucial as the DM interaction in explaining several microscopic effects like an energy gap (\(\mathrm{\sim5\ meV}\)) observed at the magnetic zone center (Fig 5a). It is already known that neither the ordinary easy axis anisotropy term (the third term in equation (1)) nor the fourth-order anisotropy term can produce the energy gap since the effect of these terms coming from each Mn site cancel out\(\mathrm{^{25,27}}\). Considering symmetrically allowed two-ion anisotropy is also unnecessary since this term would destroy the subtle spin canting observed experimentally\(\mathrm{^{13}}\) (see Supplementary Note 4 for more details.) Thus we applied the sixth-order anisotropy term (the fourth term in equation (1)), and calculated spin waves near the magnetic zone center (Fig 5c). For the details of this calculation, see Supplementary Note 4. The calculation result clearly reproduces the energy gap at the C point (Fig 5c).

Moreover, the single ion anisotropy term related to the easy-axes (the third term in equation (1)) can explain the anharmonicity of the helical order. As noted before, a magnetic Bragg peak also exists at (1,0,3\(\mathrm{\tau_{2}}\)) in addition to the primary satellite peak at (1,0,\(\mathrm{\tau_{2}}\))\(\mathrm{^{17}}\). In terms of Fourier components, this can be interpreted as the third order anharmonicity of the helical ordering; some part of which shows a tendency to rotate three times the normal angle per unit cell. However, finding a corresponding magnetic structure analytically is very difficult. Instead, we performed a Monte Carlo simulation with our Hamiltonian to obtain the energy-minimized magnetic structure, and calculated the structure factor along the (1 0 L) line (See Methods). Interestingly enough, the simulation results show that the easy-axis anisotropy term clearly produces the third order anharmonicity (Fig 5e), which is also consistent with the data from the previous study (ref. 17).

\section*{Discussion}
It is worth pointing out the key difference between our data and the previous reported results, extending the comparison with our experimental results. Supported by our DFT calculations, taking exchange interactions up to the \(\mathrm{3^{rd}}\) nearest neighbor into account in our Hamiltonian is consistent with that in the previous INS study of \(\mathrm{Mn_{3}Sn}\) (ref. 25). However, the corresponding values do differ; especially for the bonds with the almost same bond-length but different symmetry. For example, \(\mathrm{J_{4}}\) should be almost half the value of \(\mathrm{J_{5}}\) to fit the data, while ref. 25 assumed them as identical. We attribute the difference between \(\mathrm{J_{4}}\) and \(\mathrm{J_{5}}\) to a different symmetry due to the different Sn environments, (See Supplementary Fig. 4) and the same feature holds also for \(\mathrm{J_{2}}\) and \(\mathrm{J_{3}}\). Our DFT calculation indeed demonstrates the difference between \(\mathrm{J_{4}}\) and \(\mathrm{J_{5}}\) as well, giving multiple values for the bonds with the same bond length (See supplementary Fig. 5). Also, our fitting parameters show similar oscillatory behavior to that of the parameters obtained from DFT calculation of \(\mathrm{Mn_{3}Ir}\) (ref. 31, 33), which has a structure comparable to that of \(\mathrm{Mn_{3}Sn}\) (See Supplementary Fig. 5). However, it is to be noted that both the parameters in Table 1 and those from the DFT calculations do not follow the RKKY exchange curve. The discrepancy between those and the RKKY-curve may come from the non-sphericity of Fermi surfaces and the multiband contributions, which are not negligible for \(\mathrm{Mn_{3}Sn}\) (ref. 34). Note also that in real metals, an RKKY-type distance dependence of the exchange interaction applies only in the asymptotic limit.  Meanwhile, our Hamiltonian is also similar to that used in the recent theoretical paper about \(\mathrm{Mn_{3}Sn}\) (ref. 30) in terms of anisotropic contributions, except that we have additionally considered \(\mathrm{6^{th}}\) order anisotropy.

Given that the intrinsic magnetic Hamiltonian of the A-type and the B-type \(\mathrm{Mn_{3}Sn}\) cannot differ greatly from each other at room temperature, and the intrinsic Hamiltonian of the A-type \(\mathrm{Mn_{3}Sn}\) does not change significantly with temperature\(\mathrm{^{22}}\) (see Supplementary Fig. 6), our work would be reasonably applicable to the intrinsic Hamiltonian of the B-type \(\mathrm{Mn_{3}Sn}\). Nevertheless, we would like to note that performing inelastic neutron scattering for the B-type \(\mathrm{Mn_{3}Sn}\) will be definitely interesting to do. The distinct difference between the two types is the amount of excess Mn atoms occupying the 2\textit{c} sites (Sn sites), which is indeed responsible for the emergence of the spin glass phase near 50 K rather than the helical ordering in the B-type \(\mathrm{Mn_{3}Sn}\)\(\mathrm{^{20}}\). When the temperature is much higher than 50 K, the effect of these excess Mn atoms is marginal, and we expect the B-type \(\mathrm{Mn_{3}Sn}\) to share much of what we found in the A-type \(\mathrm{Mn_{3}Sn}\) since the helical ordering through the c-axis has only a small effect on the in-plane excitation spectra. When temperature is getting closer to 50 K, however, the whole excitation spectra will change due to the different spin configuration in the glass phase of the B-type \(\mathrm{Mn_{3}Sn}\). Also, spin-wave stiffness and magnon lifetime would decrease, and some quasi-elastic peaks may appear, as observed in \(\mathrm{Fe_{x}Cr_{1-x}}\) alloy system\(\mathrm{^{35}}\) where the glass phase emerges under the situation similar to \(\mathrm{Mn_{3}Sn}\). Yet this effect should be less dramatic in \(\mathrm{Mn_{3}Sn}\), since it partially retains the ferromagnetic long-range order in the glass phase.

The change of Weyl points and anomalous Hall conductivity in the low temperature phase with the helical ordering is quite an interesting point. Surprisingly, a recent study has shown that the anomalous Hall effect from the Berry curvature of Weyl points almost disappears in the low-T helical ordering phase of the A-type \(\mathrm{Mn_{3}Sn}\) (ref. 21). The simplest way to explain this change is to look at the symmetry of the magnetic structure. Due to the presence of nearly exact two-fold screw symmetry with 5.5c fractional translation along the c-axis, the anomalous Hall conductivity (\(\mathrm{\sigma_{xz}}\) and \(\mathrm{\sigma_{yz}}\)) should be almost zero, consistent with ref. 21. Of further interest, the helical ordering also breaks the inversion symmetry of the magnetic structure. If both inversion and time-reversal symmetries are broken, Weyl points may disappear by pair annihilation or they might still exist with changed position\(\mathrm{^{36,37}}\). To examine the effect of the helical ordering on Weyl points, we performed a simple band calculation using the tight binding method (See Supplementary Fig. 7), following the theoretical model initially proposed for the stacked kagome lattice\(\mathrm{^{38}}\), which is equivalent to \(\mathrm{Mn_{3}Sn}\). The results clearly show that Weyl points disappear under the helical ordering, mainly due to both significant zone-folding and spin-orbit coupling. Thus, although this calculation is too simple to be compared in full detail to the real band structure of \(\mathrm{Mn_{3}Sn}\), at least our result shows that the helical ordering is generally a sufficient perturbation to eliminate Weyl points in the electronic band structure and therefore cause the anomalous Hall effect to decrease significantly for the A-type \(\mathrm{Mn_{3}Sn}\). 

To summarize, we report the spin waves measured over the full Brillouin zone for the intermetallic antiferromagnet A-type \(\mathrm{Mn_{3}Sn}\), and present the full spin Hamiltonian. Further supported by DFT calculations, our Hamiltonian produces the helical order with \(k_z = 0.0904\) as a magnetic ground state, including subtle anharmonicity. Our subsequent analysis supported by theoretical calculations using a tight binding model reveals that this helical ordering of the A-type \(\mathrm{Mn_{3}Sn}\) removes the Weyl point, and consequently reduces the anomalous Hall conductivity experimentally seen in the B-type \(\mathrm{Mn_{3}Sn}\). Our work offers a rare example of how a topological phenomenon, namely the Weyl point, is removed by introducing a subtle change to the magnetic ground state. Ultimately, it demonstrates an intricate coupling between the electronic and spin degrees of freedom in inducing the exotic topological phase in the B-type \(\mathrm{Mn_{3}Sn}\).

\newenvironment{methods}{%
    \section*{Methods}%
    \setlength{\parskip}{12pt}%
    }{}
    
\begin{methods}
\small
\doublespacing
\subsection{Sample preparation.}
The \(\mathrm{Mn_{3}Sn}\) single crystal with mass \(\mathrm{\sim} \) 6 g was prepared by the Bridgman method. Pure Mn (99.98\%) and Sn (99.999\%) pieces in a molar ratio 3:1.1 were placed into a point bottom-shaped alumina crucible (99.8\%) and sealed in quartz ampoules under 0.3 bar argon atmosphere (99.9999\%). The material was first pre-reacted in a box furnace and then placed in a custom-made Bridgeman furnace with vertical temperature gradient. Starting in the hot zone at 1000 \textcelsius\ , the sample was slowly moved to the colder bottom zone of the furnace for 4 days. After this process, the ingot was characterized by Laue diffraction and scanning electron microscopy (MIRA, Tescan Cezch Republic) equipped with an EDX detector (XFlash, Bruker Germany). A large single crystal of \(\mathrm{Mn_{3}Sn}\) was obtained from the bottom part of the ingot, while intergrowths of \(\mathrm{Mn_{3}Sn_{2}}\) were found in the top part; this part of the ingot was removed. Before further measurements, the high quality of the crystal was then confirmed again with an IP-XRD Laue Camera (TRY-IP-YGR, TRY SE). The precise stoichiometric ratio between Mn and Sn was analyzed using an ICP-Atomic Emission Spectrometer (OPTIMA 8300, Perkin-Elmer USA), giving a result of \(\mathrm{Mn_{2.983}Sn_{1.017}}\). Magnetization was also measured to check the bulk properties (MPMS-3 EverCool, Quantum Design USA). The results were consistent with the previously reported results (See Supplementary Fig. 1).

\subsection{Inelastic neutron scattering experiment.}
Inelastic neutron scattering experiment was done on the single crystal using the MERLIN time-of-flight spectrometer at ISIS, UK\(\mathrm{^{39, 40}}\). We used highly pure Al sample holder with proper Cd shielding to minimize unnecessary background signals. The measurements were performed at 5 K. By using the repetition-rate-multiplication (RRM) technique\(\mathrm{^{41}}\), we could collect data simultaneously with various incident neutron energies (23, 42 and 100 meV, with chopper frequency of 400 Hz) in order to obtain the wide energy-momentum spectra with high quality. To cover the full spin wave spectra, the measurement with an incident neutron energy of 200 meV was done using a sloppy chopper (with chopper frequency of 500 Hz) in a single \(\mathrm{E_{i}}\) mode. Following standard procedures, time-independent background was removed from the raw data using the data collected between 15000 μs and 18000 μs, with the data corrections to compensate for the detector efficiency\(\mathrm{^{42}}\). All data were symmetrized while conserving the lattice symmetry to improve statistics. Then we properly combined all the data from different neutron incident energies to obtain uniformly fine quality over a wide energy and momentum range. After the experiment, we analyzed the data using the Horace software\(\mathrm{^{43}}\).

\subsection{Theoretical calculations.}
We used spinW to calculate theoretical spin wave spectra, which is based on a linear spin wave theory using the method of diagonalizing the magnetic Hamiltonian by applying the Holstein-Primakoff transformation after local coordinate rotation\(\mathrm{^{44}}\). For the simulation showing the emergence of anharmonicity in the helical ordering, we set a supercell of [1 1 6600] size and executed the energy-minimizing algorithm built in spinW to obtain the corresponding magnetic structure. Omitting some constants, the following well-known equation is used to calculate the structure factor.
\begin{equation}
\frac{d\mathrm{\sigma}}{d\mathrm{\Omega}} =(F(q))^{2}\sum_{ij}^{}{(\delta_{ij}- \hat{q_{i}} \hat{q_{j}})}\sum_{\lambda_{k}}^{} {p_{\lambda_{k}}} \times \sum_{l,l'}^{} {\mathrm{exp}(i\mathbf{q}\cdot(\mathbf{r}_{l'}-\mathbf{r}_{l}))\langle \lambda_{k}|s^{i}_{l}s^{j}_{l'}|\lambda_{k}\rangle}
\end{equation}
where F(q) and \(\mathrm{p_{\lambda_{k}}}\) denotes the magnetic form factor and polarization, respectively.
The damping effect is also included by an additional convolution with a Lorentzian function in addition to the Gaussian convolution related to the limited instrumental resolution. The instrumental resolution was precisely determined by the MantidPlot program, with accurate consideration of the beamline's specifications and chopper frequency\(\mathrm{^{45}}\). For the Lorentzian convolution, we assumed that the amount of damping depends only on magnon energy. Dependence of its half width at half maximum (\Gamma) on spin wave energy (E) was chosen as the following, where E is a variable with a step of 0.1 meV: \Gamma=1 meV for energies less than 40 meV, \Gamma=E/64 for E = 40.1 to 85 meV, and \Gamma=E/26 for E = 85.1 to 110 meV. Those values are determined by fitting the calculated intensity with the data.

Scalar-relativistic ab-initio calculations were carried out using a Green's function approach based on the linearized muffin-tin orbital (LMTO) method in the atomic sphere approximation\(\mathrm{^{46}}\), as implemented in the Questaal code\(\mathrm{^{47}}\). To compensate for the underestimation of the exchange splitting in itinerant Mn compounds typical for the local density approximation (LDA), we employed the LDA+U method with double counting correction in the fully-local limit\(\mathrm{^{48}}\) using the parameters U=1.0 eV and J=0.5 eV. The isotropic exchange interactions were then calculated in terms of the method of infinitesimal rotations\(\mathrm{^{49}}\) from a ferromagnetic state of reference. The Dzyaloshinskii-Moriya interactions were obtained by exploiting the spin-cluster expansion technique as implemented in combination with the relativistic scheme of Disordered Local Moments within the screened Korringa-Kohn-Rostoker Green's function approach\(\mathrm{^{30}}\).

\subsection{Data availability.}
All relevant data that support the findings of this study are available from the corresponding author on request.
\end{methods}

\newenvironment{addendum}{%
    \setlength{\parindent}{0in}%
    \small%
    \begin{list}{Acknowledgements}{%
        \setlength{\leftmargin}{0in}%
        \setlength{\listparindent}{0in}%
        \setlength{\labelsep}{0em}%
        \setlength{\labelwidth}{0in}%
        \setlength{\itemsep}{12pt}%
        \let\makelabel\addendumlabel}
    }
    {\end{list}\normalsize}

\newcommand*{\addendumlabel}[1]{\textbf{#1}\hspace{1em}}

\begin{addendum}
\small
\setlength{\baselineskip}{6mm}

  \item We thank Kisoo Park, Jon Leiner, Taehun Kim and L. Udvardi for helpful discussions, and M. Valiska and C. Draser for technical assistance. The work at the IBS CCES was supported by the research programme of Institute for Basic Science (IBS-R009-G1). The work at the Materials Growth and Measurement Laboratory MGML was supported by the grant no. LM2011025 of the Ministry of Education, Youth and sports of Czech Republic. A.D. and L.S. acknowledge support provided by the BME-Nanotechnology FIKP grant of EMMI (BME FIKP-NAT) and by the Hungarian National Scientific Research Fund (NKFIH) under projects no.K115575 and PD124380. J.J. acknowledges support from the UK’s EPSRC under a service level agreement with STFC (core support for the CCP9 project).

 \item[Competing interests]The authors declare no competing financial interests.

 \item[Author contributions] J.G.P. initiated and supervised the project. K.U. and V.S. grew the single crystal and J.O., H.K., H.C. and P.P. measured bulk properties. J.O., K.U., H.C.W. and D.A. conducted the inelastic neutron scattering experiments. P.P. and J.O. made the spin waves calculations, and P.P. \& K.H.L. did tight binding calculations. DFT calculations were carried out by J.J., L.S., and A.D. All authors contributed to the discussion and writing the paper.
\end{addendum}

\small
\setlength{\parindent}{0em}
\textbf{Supplementary Information} accompanies the paper on the \textit{npj Quantum Materials} website.

\section*{References}
\small
\begin{enumerate}[$\hspace{0pt} 1. $]
\doublespacing
\itemsep1.2mm 
\item Hasan, M. Z. \& Kane, C. L. Colloquium\(\colon\)Topological insulators. \textit{Rev. Mod. Phys} \textbf{82}, 3045-3067 (2010).

\item Kübler, J. \& Felser, C. Non-collinear antiferromagnets and the anomalous Hall effect. \textit{EPL (Europhysics Letters)} \textbf{108}, 67001 (2014).

\item Chen, H., Niu, Q. \& MacDonald, A. H. Anomalous Hall effect arising from noncollinear antiferromagnetism. \textit{Phys. Rev. Lett.} \textbf{112}, 017205 (2014).

\item Nakatsuji, S., Kiyohara, N. \& Higo, T. Large anomalous Hall effect in a non-collinear antiferromagnet at room temperature. \textit{Nature} \textbf{527}, 212-215 (2015).

\item Zhang, Y. et al. Strong anisotropic anomalous Hall effect and spin Hall effect in the chiral antiferromagnetic compounds ${\mathrm{Mn}}_{3}X$ ($X=\mathrm{Ge}$, Sn, Ga, Ir, Rh, and Pt). \textit{Phys. Rev. B} \textbf{95}, 075128 (2017).

\item Ikhlas, M. et al. Large anomalous Nernst effect at room temperature in a chiral antiferromagnet. \textit{Nat. Phys.} \textbf{13}, 1085-1090 (2017).

\item Li, X. et al. Anomalous Nernst and Righi-Leduc effects in ${\mathrm{Mn}}_{3}\mathrm{Sn}$: Berry curvature and entropy flow. \textit{Phys. Rev. Lett.} \textbf{119}, 056601 (2017).

\item Guo, G.-Y. \& Wang, T.-C. Large anomalous Nernst and spin Nernst effects in the noncollinear antiferromagnets ${\mathrm{Mn}}_{3}X$ ($X=\mathrm{Sn},\mathrm{Ge},\mathrm{Ga}$). \textit{Phys. Rev. B} \textbf{96}, 224415 (2017).

\item Kuroda, K. et al. Evidence for magnetic Weyl fermions in a correlated metal. \textit{Nature Materials} \textbf{16}, 1090-1095 (2017).

\item Zelezny, J., Zhang, Y., Felser, C. \& Yan, B. Spin-polarized current in noncollinear antiferromagnets. \textit{Phys. Rev. Lett.} \textbf{119}, 187204 (2017).

\item Tomiyoshi, S. \& Yamaguchi, Y. Magnetic structure and weak ferromagnetism of \(\mathrm{Mn_{3}Sn}\) studied by polarized neutron diffraction. \textit{J. Phys. Soc. Jpn} \textbf{51}, 2478-2486 (1982).

\item Sticht, J., Höck, K. H. \& Kübler, J. Non-collinear itinerant magnetism: the case of \(\mathrm{Mn_{3}Sn}\). \textit{Journal of Physics: Condensed Matter} \textbf{1}, 8155-8170 (1989).

\item Nagamiya, T., Tomiyoshi, S. \& Yamaguchi, Y. Triangular spin configuration and weak ferromagnetism of \(\mathrm{Mn_{3}Sn}\) and \(\mathrm{Mn_{3}Ge}\). \textit{Solid State Communications} \textbf{42}, 385-388 (1982).

\item Brown, P. J., Nunez, V., Tasset, F., Forsyth, J. B. \& Radhakrishna, P. Determination of the magnetic structure of \(\mathrm{Mn_{3}Sn}\) using generalized neutron polarization analysis. \textit{Journal of Physics: Condensed Matter} \textbf{2}, 9409-9422 (1990).

\item Sandratskii, L. M. \& Kübler, J. Role of orbital polarization in weak ferromagnetism. \textit{Phys. Rev. Lett.} \textbf{76}, 4963-4966 (1996).

\item Krén, E., Paitz, J., Zimmer, G. \& Zsoldos, É. Study of the magnetic phase transformation in the \(\mathrm{Mn_{3}Sn}\) phase. \textit{Physica B+C} \textbf{80}, 226-230 (1975).

\item Cable, J. W., Wakabayashi, N. \& Radhakrishna, P. A neutron study of the magnetic structure of \(\mathrm{Mn_{3}Sn}\). \textit{Solid State Communications} \textbf{88}, 161-166 (1993).

\item Duan, T. F. et al. Magnetic anisotropy of single-crystalline \(\mathrm{Mn_{3}Sn}\) in triangular and helix-phase states. \textit{Applied Physics Letters} \textbf{107}, 082403 (2015).

\item Ohmori, H., Tomiyoshi, S., Yamauchi, H. \& Yamamoto, H. Spin structure and weak ferromagnetism of \(\mathrm{Mn_{3}Sn}\). \textit{Journal of Magnetism and Magnetic Materials} \textbf{70}, 249-251 (1987).

\item Feng, W. J. et al. Glassy ferromagnetism in \(\mathrm{Ni_{3}Sn}\)-type \(\mathrm{Mn_{3.1}Sn_{0.9}}\). \textit{Phys. Rev. B} \textbf{73}, 205105 (2006).

\item Sung, N. H., Ronning, F., Thompson, J. D. \& Bauer, E. D. Magnetic phase dependence of the anomalous Hall effect in \(\mathrm{Mn_{3}Sn}\) single crystals. \textit{Applied Physics Letters} \textbf{112}, 132406 (2018).

\item Zimmer, G. J. \& Kren, E. Investigation of the magnetic phase transformation in \(\mathrm{Mn_{3}Sn}\). \textit{AIP Conference Proceedings} \textbf{5}, 513-516 (1972).

\item Radhakrishna, P. \& Tomiyoshi, S. A neutron scattering study of the magnetic excitations in a triangular itinerant antiferromagnet, \(\mathrm{Mn_{3}Sn}\). \textit{Journal of Physics: Condensed Matter} \textbf{3}, 2523-2527 (1991).

\item Radhakrishna, P. \& Cable, J. W. Magnetic excitations in the triangular antiferromagnet \(\mathrm{Mn_{3}Sn}\). \textit{Journal of Magnetism and Magnetic Materials} \textbf{104-107}, 1065-1066 (1992).

\item Cable, J. W., Wakabayashi, N. \& Radhakrishna, P. Magnetic excitations in the triangular antiferromagnets \(\mathrm{Mn_{3}Sn}\) and \(\mathrm{Mn_{3}Ge}\). \textit{Phys. Rev. B} \textbf{48}, 6159-6166 (1993).

\item Cable, J.W., Wakabayashi, N. \& Radhakrishna, P. Structure of the modulated magnetic phase of \(\mathrm{Mn_{3}Sn}\). United States. doi: https://www.osti.gov/servlets/purl/10104409 (1993).

\item J. Jensen \& A. R. Mackintosh, \textit{Rare-Earth Magnetism}, Ch. 5 (Clarendon, Oxford, 1991)

\item Diallo, S. O. et al. Itinerant magnetic excitations in antiferromagnetic ${\mathrm{CaFe}}_{2}{\mathrm{As}}_{2}$. \textit{Phys. Rev. Lett.} \textbf{102}, 187206 (2009).

\item Zhao, J. et al. Spin waves and magnetic exchange interactions in ${\mathrm{CaFe}}_{2}{\mathrm{As}}_{2}$. \textit{Nat. Phys.} \textbf{5}, 555-560 (2009).

\item Liu, J. \& Balents, L. Anomalous Hall effect and topological defects in antiferromagnetic Weyl semimetals: \(\mathrm{Mn_{3}Sn/Ge}\). \textit{Phys. Rev. Lett.} \textbf{119}, 087202 (2017).

\item Szunyogh, L., Udvardi, L., Jackson, J., Nowak, U. \& Chantrell, R. Atomistic spin model based on a spin-cluster expansion technique: Application to the \(\mathrm{IrMn_{3}/Co}\) interface. \textit{Phys. Rev. B} \textbf{83}, 204401 (2011).

\item Jeong, J. et al. Temperature-dependent interplay of Dzyaloshinskii-Moriya interaction and single-ion anisotropy in multiferroic \(\mathrm{BiFeO_{3}}\). \textit{Phys. Rev. Lett.} \textbf{113}, 107202 (2014).

\item Szunyogh, L., Lazarovits, B., Udvardi, L., Jackson, J. \& Nowak, U. Giant magnetic anisotropy of the bulk antiferromagnets \(\mathrm{IrMn}\) and \(\mathrm{IrMn_{3}}\) from first principles. \textit{Phys. Rev. B} \textbf{79}, 020403 (2009).

\item Alireza, A., Peter, T. \& Ilya, E. Evolution of the multiband Ruderman–Kittel–Kasuya–Yosida interaction\(\colon\) application to iron pnictides and chalcogenides. \textit{New Journal of Physics} \textbf{15}, 033034 (2013).

\item Fincher, C. R., Shapiro, S. M., Palumbo, A. H. \& Parks, R. D. Spin-wave evolution crossing from the ferromagnetic to spin-glass regime of ${\mathrm{Fe}}_{x}{\mathrm{Cr}}_{1\ensuremath{-}x}$. \textit{Phys. Rev. Lett.} \textbf{45}, 474-477 (1980).

\item Zyuzin, A. A., Wu, S. \& Burkov, A. A. Weyl semimetal with broken time reversal and inversion symmetries. \textit{Phys. Rev. B} \textbf{85}, 165110 (2012).

\item Chang, G. et al. Magnetic and noncentrosymmetric Weyl fermion semimetals in the 
RAlGe family of compounds (R=rare earth). \textit{Phys. Rev. B} \textbf{97}, 041104 (2018).

\item Ito, N. \& Nomura, K. Anomalous Hall effect and spontaneous orbital magnetization in antiferromagnetic Weyl metal. \textit{J. Phys. Soc. Jpn} \textbf{86}, 063703 (2017).

\item Bewley, R., Guidi, T. \& Bennington, S. MERLIN: a high count rate chopper spectrometer at ISIS. \textit{Notiziario Neutroni e Luce di Sincrotrone} \textbf{14}, 22-27 (2009).

\item Park, J.-G. et al; (2016): 1600054, \textit{STFC ISIS Facility}, doi:10.5286/ISIS.E.79298815.

\item Russina, M. \& Mezei, F. First implementation of repetition rate multiplication in neutron spectroscopy. \textit{Nuclear Instruments and Methods in Physics Research Section A: Accelerators, Spectrometers, Detectors and Associated Equipment} \textbf{604}, 624-631 (2009).

\item C. G. Windsor, \textit{Pulsed Neutron Scattering} (Taylor \& Francis Ltd, London, 1981)

\item Ewings, R. A. et al. Horace: Software for the analysis of data from single crystal spectroscopy experiments at time-of-flight neutron instruments. \textit{Nuclear Instruments and Methods in Physics Research Section A: Accelerators, Spectrometers, Detectors and Associated Equipment} \textbf{834}, 132-142 (2016).

\item Toth, S. \& Lake, B. Linear spin wave theory for single-Q incommensurate magnetic structures. \textit{J. Phys. Condens. Matter} \textbf{27}, 166002 (2015).

\item Arnold, O. et al. Mantid—Data analysis and visualization package for neutron scattering and μSR experiments. \textit{Nuclear Instruments and Methods in Physics Research Section A: Accelerators, Spectrometers, Detectors and Associated Equipment} \textbf{764}, 156-166 (2014).

\item Andersen, O. K. \& Jepsen, O. Explicit, first-principles tight-binding theory. \textit{Phys. Rev. Lett.} \textbf{53}, 2571-2574 (1984).

\item Questaal Electronic Structure Package, https://www.questaal.org

\item Liechtenstein, A I,  Anisimov, V I \& Zaanen, J. Density-functional theory and strong interactions: Orbital ordering in Mott-Hubbard insulators. \textit{Phys. Rev. B} \textbf{52}, R5467-R5470 (1995).

\item Liechtenstein, A. I., Katsnelson, M. I., Antropov, V. P. \& Gubanov, V. A. Local spin density functional approach to the theory of exchange interactions in ferromagnetic metals and alloys. \textit{Journal of Magnetism and Magnetic Materials} \textbf{67}, 65-74 (1987). 
\end{enumerate}

\newpage
\begin{figure}
\setlength{\baselineskip}{4mm}
\includegraphics[width=0.9\textwidth]{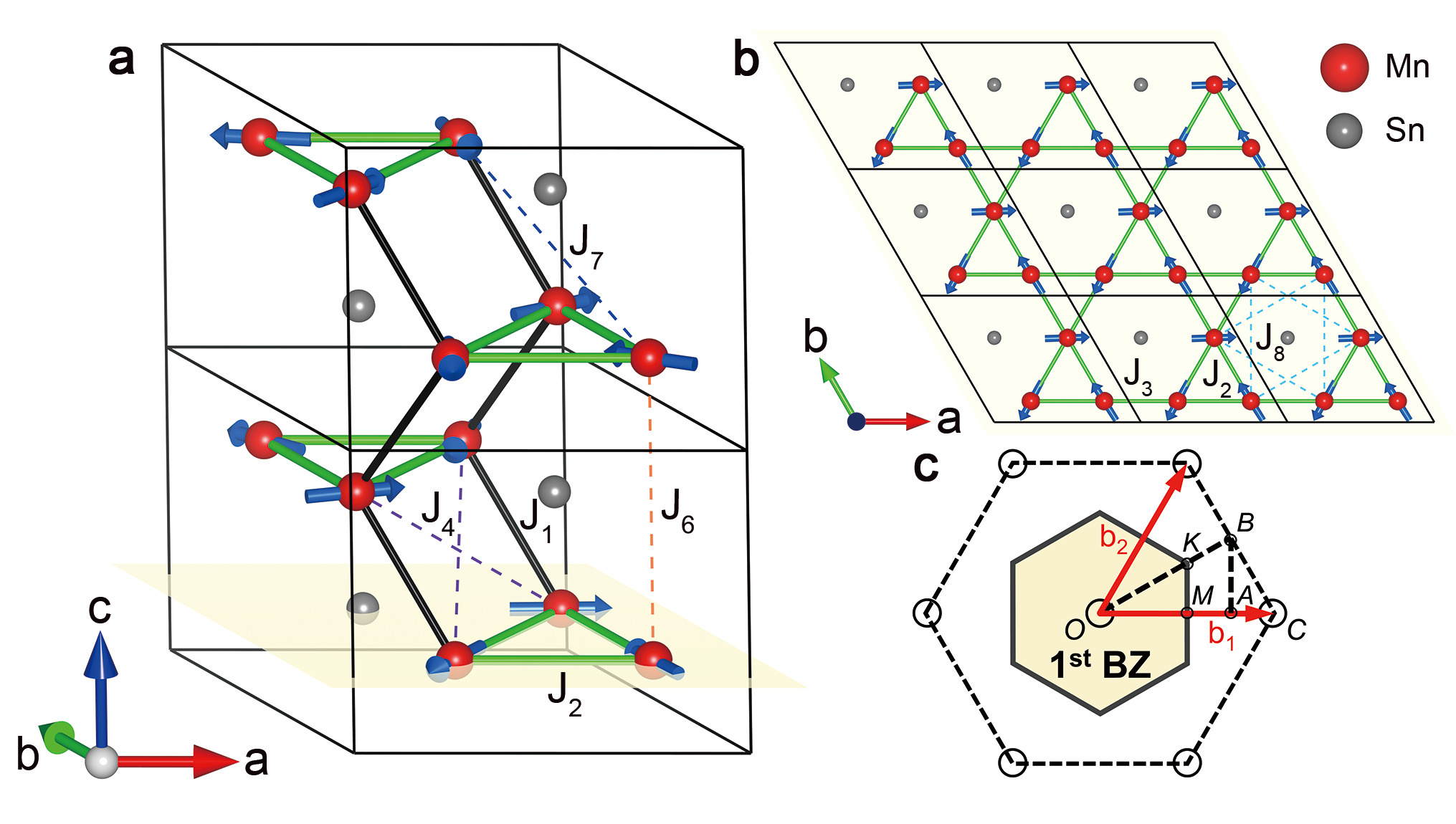} 
\centering
\caption{\textbar\textbf{ Magnetic structure and reciprocal lattice of \boldmath\(\mathsf{Mn_{3}Sn}\).} (\textbf{a}) Magnetic structure of \(\mathsf{Mn_{3}Sn}\) with helical ordering and possible paths for exchange interactions. Lattice parameters are a=b=5.67 \AA, and c=4.53 \AA. Note that inversion symmetry is broken in the real magnetic structure. For example, two spins connected by \(\mathsf{J_{4}}\)
are not parallel, which makes the spins non-centrosymmetric. In addition, see Supplementary Fig. 4 for \(\mathsf{J_{5}}\) which extends to an adjacent unit cell and so cannot be drawn in this picture. (\textbf{b}) An a-b plane formed by Mn and Sn atoms at z=0.25 (l.u), which corresponds to a yellow a-b plane in \textbf{a}. Nearest in-plane couplings \(\mathsf{J_{2}}\) and \(\mathsf{J_{3}}\) form a kagome shape. In fact, the structure is very slightly distorted from a kagome lattice, leading to a small difference in the \(\mathsf{J_{2}}\) and \(\mathsf{J_{3}}\) bond lengths (see Table 1). Note, however, that even in a perfect kagome structure the local Sn environment in adjacent layers would distinguish the \(\mathsf{J_{2}}\) and \(\mathsf{J_{3}}\) bonds. (\textbf{c}) Reciprocal lattice of \(\mathsf{Mn_{3}Sn}\) and labels of high symmetric points used in this paper.}
\end{figure}
\vskip 0.7in

\begin{figure}
\includegraphics[width=0.86\textwidth]{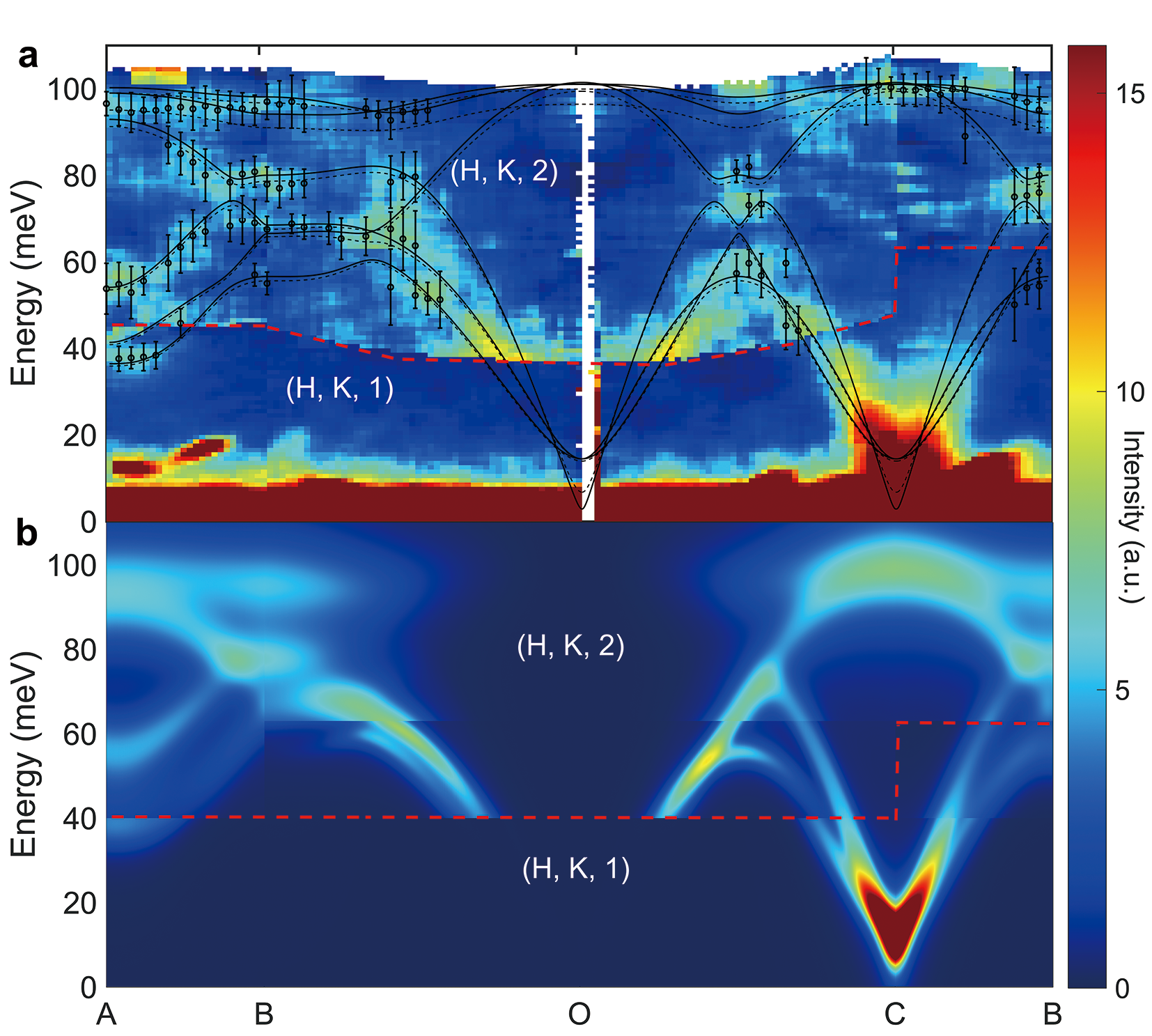} 
\centering
\setlength{\baselineskip}{4mm}
\caption{\textbar\textbf{ INS data and calculation result.} (\textbf{a}) The inelastic neutron scattering data taken at 5 K (\(\mathsf{E_{i}}\)=100 and 200 meV) and fitted dispersion (black lines) without a twin effect. Black dashed lines indicate additional magnon modes coming from the incommensurate helical order. As indicated with red dashed lines, we combined the data from both (HK1) and (HK2), since the data at (HK2) plane have a wider energy coverage; for the low energy part (\textless40 meV) these data are, however, heavily dominated by strong phonon branches and cannot be used. Note that some strong scattering below 20 meV (like between A and B) arises from multiple-scattering effects. Error bars represent one sigma of energy values, which were determined by Gaussian peak fitting of magnon modes for each constant Q-cut. (\textbf{b}) Calculated dynamical structure factors from our Hamiltonian with a magnon damping effect applied. The discontinuity along the B-O-C line and near at 60 meV arises from the data convolution with different instrumental energy resolutions (See Supplementary Fig.3 for more details).}
\end{figure}
\vskip 0.7in

\begin{figure}
\includegraphics[width=0.62\textwidth]{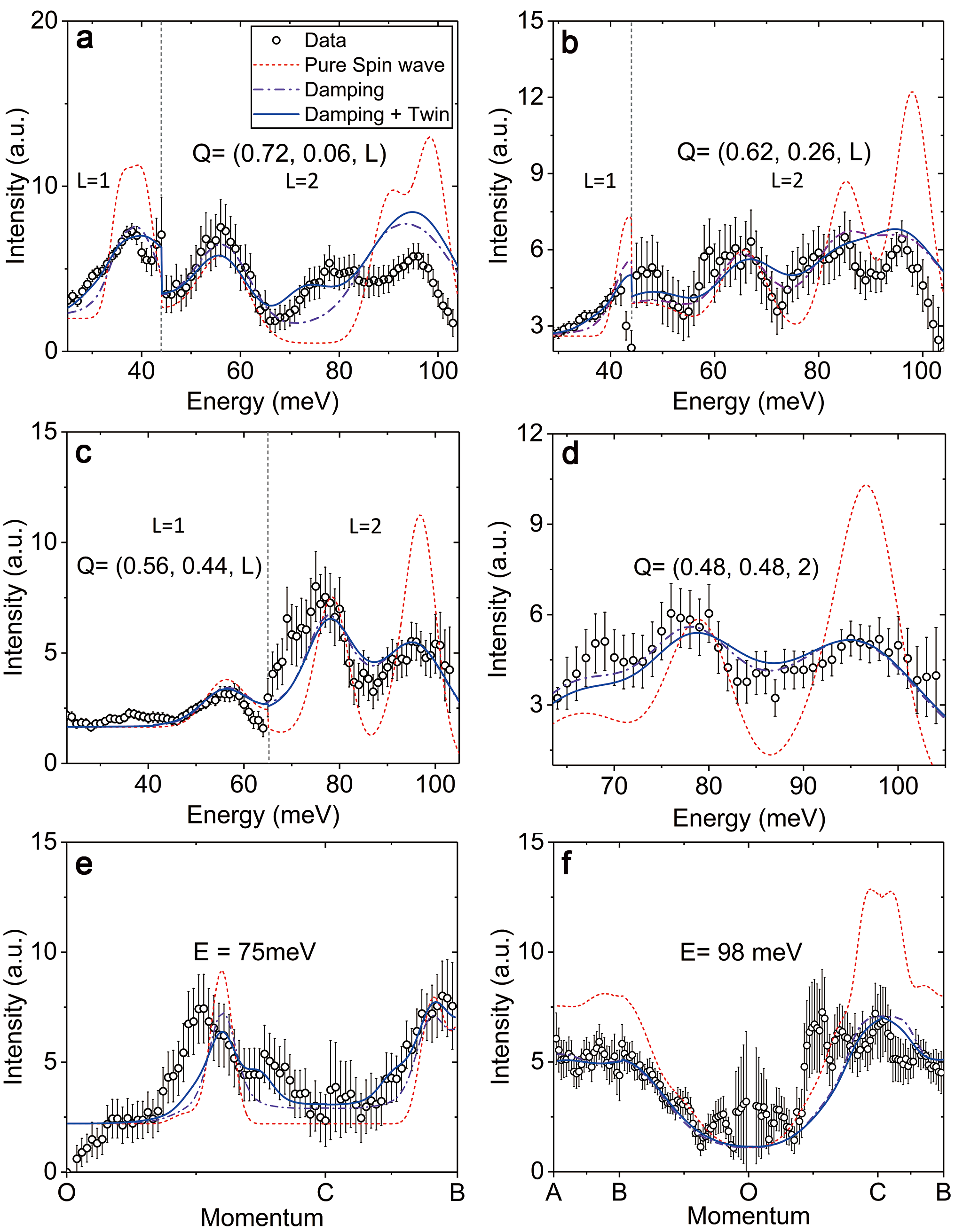} 
\centering
\setlength{\baselineskip}{4mm}
\caption{\textbar\textbf{ Comparison between data and several calculations through constant Q-cut (a-d) and constant E-cut (e,f).} Red dashed lines indicate the calculation result from a linear spin wave theory, purple dashed and dotted lines show the result with the effect of damping applied, and blue solid lines show the result considering both damping and twin effects. Gray dashed lines in \textbf{a-c} indicate a boundary between \textsf{\(\mathsf{k_{z}=1}\)} region and \textsf{\(\mathsf{k_{z}=2}\)} region. All figures clearly show the limitations of a linear spin wave theory and therefore the need for magnon damping. Calculation results including both damping and twin effects best explain the data. Note that all Q points are written in reciprocal lattice units.}
\end{figure}
\vskip 0.7in

\begin{figure}
\includegraphics[width=0.9\textwidth]{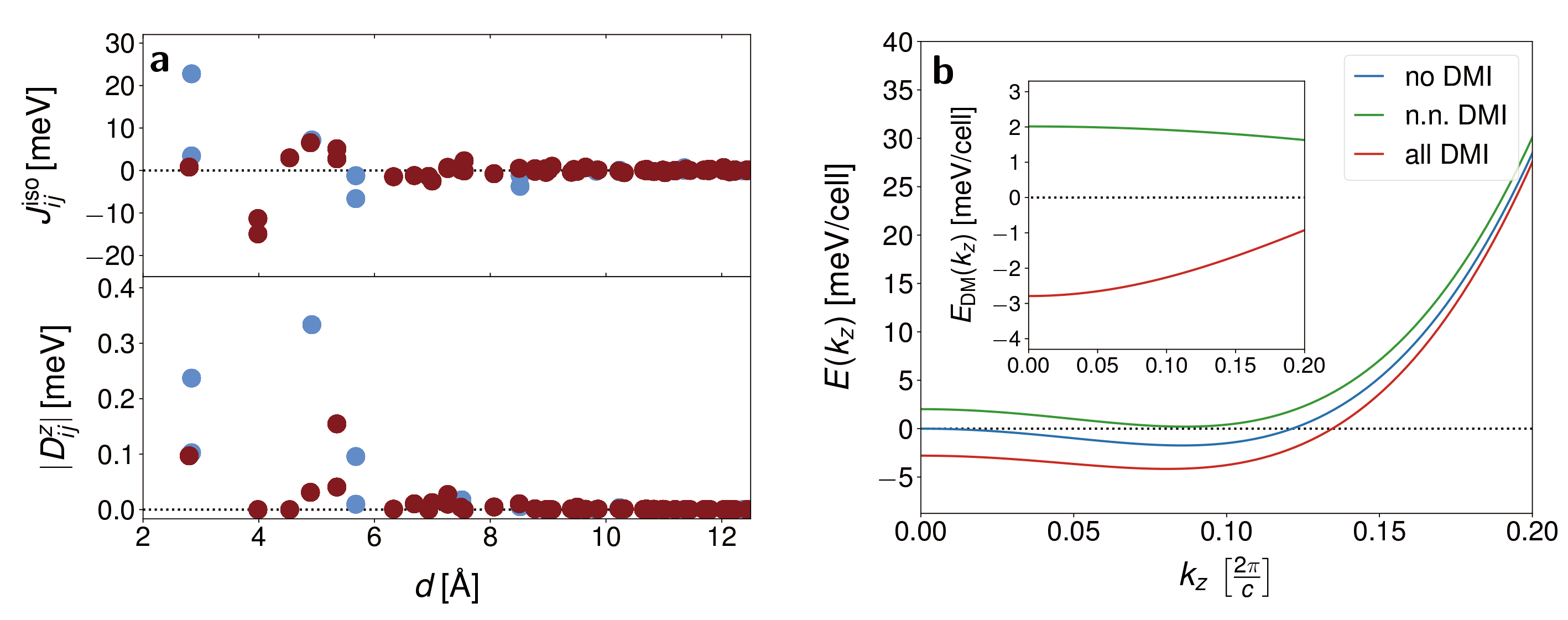} 
\centering
\setlength{\baselineskip}{4mm}
\caption{\textbar\textbf{ Spin-model parameters and stabilization of helical ordering based on ab-initio calculations.} (\textbf{a}) Isotropic Heisenberg interactions (top) and magnitude of the z-component of Dzyaloshinskii-Moriya vectors (bottom) as a function of inter-atomic distance. Red and blue circles denote inter- and intra- plane interactions, respectively. (\textbf{b}) Energy per unit cell of the (\(\mathsf{\Gamma_{5}}\)) state with a helical modulation propagating along the c direction as computed with the spin model parameters in \textbf{a}. The blue curve corresponds to a purely isotropic model, the green curve to a spin-model including also first nearest-neighbour DM interactions and the red curve refers to including all calculated DM interactions. The inset shows the contribution from the DM interactions to the spin-spiral energy for both cases. The nearest-neighbour case (the green line) fits perfectly to a function \(\mathsf{A cos(k_{z}c/2)}\).}
\end{figure}
\vskip 0.7in

\begin{figure}
\includegraphics[width=0.9\textwidth]{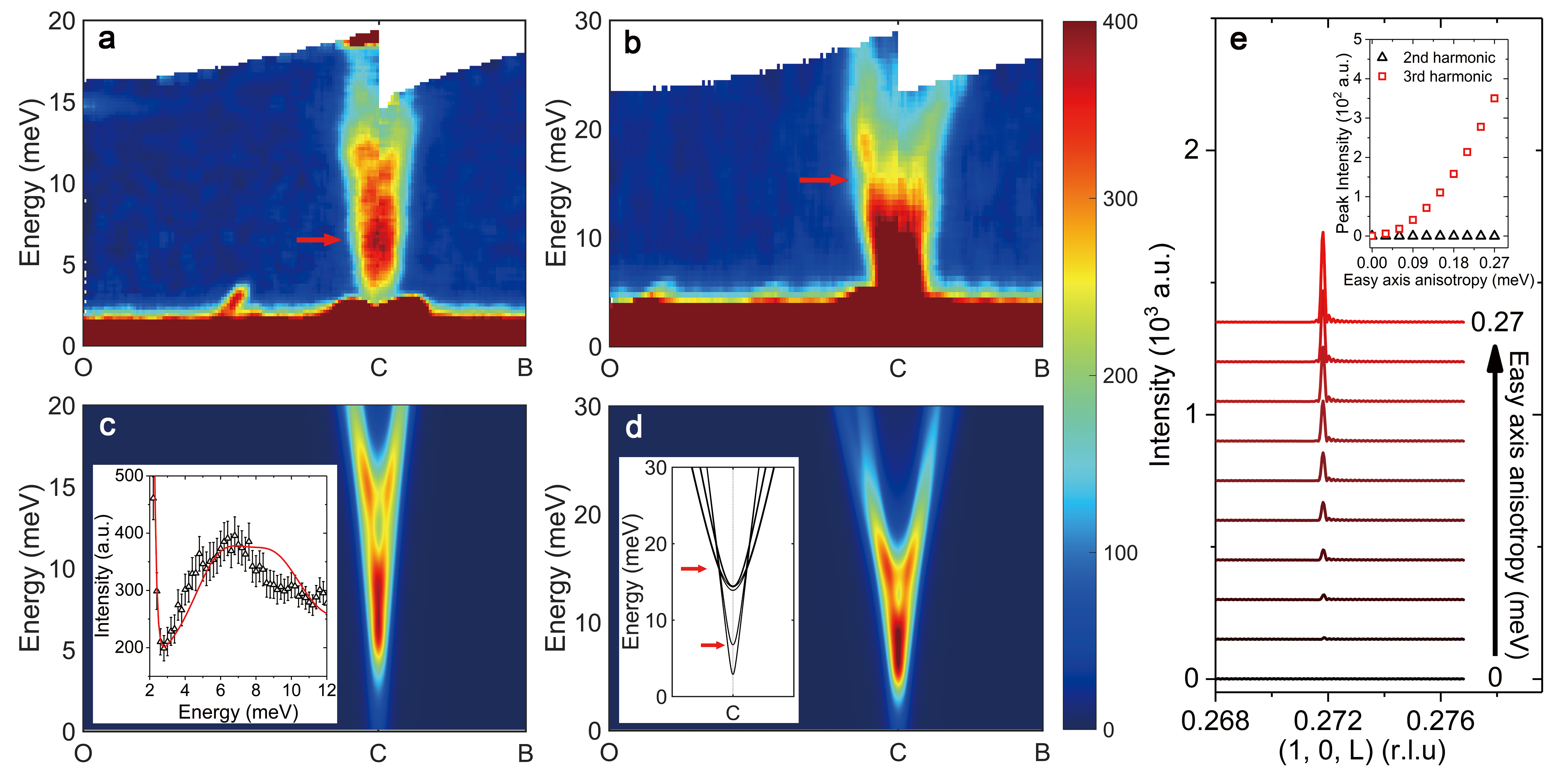} 
\centering
\setlength{\baselineskip}{4mm}
\caption{\textbar\textbf{ Low energy dynamics, and anharmonicity of the helical order.} (\textbf{a-b}) Low energy part of the inelastic neutron scattering data taken at 5 K with incident neutron beam energy \(\mathsf{E_{i}= 23\ meV}\) for \textbf{a}, and \(\mathsf{E_{i} = 42\ meV}\) for \textbf{b}. The red arrows in \textbf{a} and \textbf{b} indicate the size of an energy gap and the emergence of gradually increasing mode, respectively. Since the size of the energy gap and the height of this gradually increasing mode at the C point are determined solely by \(\mathsf{B^{6}_{6}}\) and the DM interaction, we can reasonably extract their size (Table 1). (\textbf{c-d}) Calculated dynamical structure factor paired with \textbf{a} and \textbf{b} for each. Constant Q-cut at C point (inset of \textbf{c}) whose red solid line denotes the calculation result including an elastic peak clearly shows the energy gap. (\textbf{e}) The result of Monte Carlo simulations with changing the size of easy axis anisotropy (K), showing the emergence of third harmonic satellite peak at (1, 0, 3q).}
\end{figure}

\newpage
\renewcommand{\arraystretch}{2}
\begin{table}
\scriptsize
\centering
\caption{Fitted parameters of bonds with each type, and a bond length of each one in \(\mathsf{Mn_{3}Sn}\)}
\medskip
\begin{tabular}{|c|c|c|c|c|c|c|c|c|c|c|c|c|c|}
 \hline
  & \(\mathsf{J_{1}}\) & \(\mathsf{J_{2}}\) & \(\mathsf{J_{3}}\) & \(\mathsf{J_{4}}\) & \(\mathsf{J_{5}}\) & \(\mathsf{J_{6}}\) & \(\mathsf{J_{7}}\)  & \(\mathsf{J_{8}}\) & \(\mathsf{D_{1}}\) & \(\mathsf{D_{2}}\) & \(\mathsf{D_{3}}\) & K & \(\mathsf{B^{6}_{6}}\) \\ [0.5ex] 
 \hline
 \textsf{Coupling const. (meV)} & 10.29 & 7.29 & 8.29 & -1.32 & -3.14 & 4.19 & 0 & 3.90 & 0 & 0.35\(\hat{z}\) & 0.39\(\hat{z}\) & \(-0.086 \) & \(\mathsf{2.6\times 10^{-4}}\) \\ 
 \hline
 \textsf{Bond length (\AA)} & 2.795 & 2.830 & 2.840 & 3.978 & 3.986 & 4.530 & 4.885 & 4.910 & 2.795 & 2.830 & 2.840 & & \\
 \hline
  \textsf{In-plane or Out-of-plane} & Out & In & In & Out & Out & Out & Out & In & Out & In & In & & \\[1ex] 
\hline
\end{tabular}
\end{table}
\newpage
\end{document}